\pdfoutput=1
\documentclass[letterpaper, 10 pt, conference]{ieeeconf}  

\IEEEoverridecommandlockouts                              

\usepackage{cite}
\usepackage{amsmath,amssymb,amsfonts}
\usepackage{algorithmic}
\usepackage{graphicx}
\usepackage{textcomp}
\usepackage[dvipsnames]{xcolor}
\usepackage{soul}
\usepackage{multirow}
\let\labelindent\relax
\usepackage{enumerate}
\usepackage{enumitem}
\usepackage{comment}
\usepackage{array}
\usepackage{subfigure}
\usepackage{makecell,tabularx, float}

\title{\LARGE \bf
A Model-based Multi-agent Framework to Enable an Agile Response to Supply Chain Disruptions*
}

\author{Mingjie Bi$^{1}$, Gongyu Chen$^{2}$, Dawn M. Tilbury$^{1,3}$, Siqian Shen$^{2}$ and Kira Barton$^{1,3}$
\thanks{*This work was funded in part by The United States National Science Foundation (NSF) grant \#CMMI-2034974.}
\thanks{$^{1}$Mingjie Bi is with the Robotics Institute, 
        University of Michigan, Ann Arbor, MI 48109, USA,
        {\tt\small mingjieb@umich.edu}}%
\thanks{$^{2}$Gongyu Chen and Siqian Shen are with the Department of Industrial and Operations Engineering, 
        University of Michigan, Ann Arbor, MI 48109, USA,
        {\tt\small \{chgongyu, siqian\}@umich.edu}}%
\thanks{$^{1,3}$Dawn M. Tilbury and Kira Barton are with the Robotics Institute and the Department of Mechanical Engineering, 
        University of Michigan, Ann Arbor, MI 48109, USA,
        {\tt\small \{tilbury, bartonkl\}@umich.edu}}%
}

\begin{document}

\maketitle
\thispagestyle{empty}
\pagestyle{empty}

\begin{abstract}

    Due to the COVID-19 pandemic, the global supply chain is disrupted at an unprecedented scale under uncertain and unknown trends of labor shortage, high material prices, and changing travel or trade regulations.
To stay competitive, enterprises desire agile and dynamic response strategies to quickly react to disruptions and recover supply-chain functions.
Although both centralized and multi-agent approaches have been studied, their implementation requires prior knowledge of disruptions and agent-rule-based reasoning.
In this paper, we introduce a model-based multi-agent framework that enables agent coordination and dynamic agent decision-making to respond to supply chain disruptions in an agile and effective manner.
Through a small-scale simulated case study, we showcase the feasibility of the proposed approach under several disruption scenarios that affect a supply chain network differently, and analyze performance trade-offs between the proposed distributed and centralized methods.


\end{abstract}

\section{introduction}
\label{sec:introduction}

The COVID-19 pandemic has tremendously disrupted the global supply chain in different aspects, such as customer demand, material supplies, manufacturing capabilities, labor shortage, etc.~\cite{rahman2021agent}.
It also highlights the complexity of modern global supply chains~\cite{giannakis2016multi}.
Supply chain entities, such as customers, suppliers, and manufacturers, suffer from slow knowledge propagation of supply-demand mismatch amid lock-downs and other disruptions.
Therefore, there is an urgent need for a dynamic response strategy that enables smart, agile, and resilient supply chains under unexpected disruptions~\cite{ben2019internet}.

In many supply chain networks, centralized models have been widely used to provide optimal solutions based on specific objectives (e.g., product flow cost), but have limited ability to adjust solutions due to disruptions in an agile manner~\cite{giannakis2016multi}. 
As the complexity and scale of supply chains increase, it becomes more difficult to effectively manage supply chain networks under disruptions using centralized methods~\cite{giannakis2016multi, baryannis2019supply}.
To improve the agility and effectiveness of supply chain networks, researchers have proposed distributed control strategies, where multiple entities in the system make decisions via communication and collaboration~\cite{xu2021will}.

Multi-agent control is a distributed method that enables intelligent decision-making in supply chains~\cite{baryannis2019supply}.
Each autonomous agent in a supply chain network either represents a physical entity, such as a supplier, or is responsible for a function, such as demand forecasting. %
Various agents communicate, make decisions based on their knowledge, and collaboratively solve supply chain problems~\cite{toorajipour2021artificial,baryannis2019supply}.

In this work, we focus on the problem of reacting to disruptions and recovering supply chain functions: given a supply chain network, existing product flows, and a disruption, we aim to identify a new product flow plan that satisfies the demand and minimizes penalties in terms of costs, time, etc.
In the existing literature, most agent-based disruption reaction strategies are based on 
pre-defined disruption scenarios and reactive actions via a stochastic programming model~\cite{torabi2015resilient}, a Petri Nets model~\cite{blos2018framework}, or a case-based disruption reaction database~\cite{giannakis2011multi}.
In many examples, the disruption reaction performance of these methods is limited to a set of pre-defined scenarios, and it is can be difficult and even impossible to cope with unexpected disruptions outside of the predefined set.

To address this limitation, agents need to be equipped with model-based knowledge to make decisions dynamically.
However, existing multi-agent approaches focus on either system-level architectures or rule-based agents~\cite{xu2021will}.
The works~\cite{rahman2021agent,fiedler2022agent,hajian2022agent} provide general descriptions of agent attributes and functions at a conceptual level. 
References~\cite{ghadimi2019intelligent, pal2014multi, wang2013ontology} use rule-based reasoning logic to guide agent decisions, while~\cite{giannakis2016multi, shukla2016fuzzy, pal2014multi} introduce case-based agents that provide pre-planned decision-making and coordinated behaviors for the agents.  
In these methods, relying on a rule-based strategy limits the ability of the agents to readily adapt to unexpected supply chain disruptions.
Additionally, it is difficult to scale this rule-based approach to larger and more complex systems, with the addition of more rules reducing the flexibility of agent behaviors.
Lastly, references~\cite{kovalenko2019model,bi2021dynamic} introduce model-based agents for manufacturing systems; however, such approaches to define model-based agents for supply chain networks are lacking.
        

The main contributions of this paper include: (1) deriving a modeling structure within the knowledge base of supply-chain agents, (2) development of a dynamic agent communication infrastructure to respond to unknown disruptions, and (3) evaluation of potential trade-offs between centralized and the proposed distributed multi-agent methods through a simulated case study with various types of disruptive events.

The rest of the paper is organized as follows. 
In Section~\ref{sec:centralizedmodel}, we introduce a centralized optimization model to optimize flow in supply chains.
In Section~\ref{sec:multiagentframework}, we present the proposed distributed multi-agent framework.
In Section~\ref{sec:disruptionresponse}, we describe an agent communication infrastructure for disruption response.
Section~\ref{sec:casestudy} provides a case study for approach comparisons, with concluding remarks given in Section~\ref{sec:conclusion}.

\section{centralized model}
\label{sec:centralizedmodel}
In this section we introduce a standardized, centralized model to serve as a comparison to the proposed multi-agent framework. In this work, we modify a traditional network-flow model that only considers product flow~\cite{ahuja1993network} and extend it to a generic mixed-integer linear programming (MILP) model that considers production, inventory, and transportation planning in response to disruptions for a multi-echelon, multi-product supply chain network.
\subsection{A centralized model for supply chain management}
\label{sec:genericmodel}

Consider a supply chain network $G(V,E)$ with $V$ being the set of vertices and $E$ being the set of edges.
The vertices represent the supply chain entities, such as suppliers, customers, etc., and the edges represent possible material or product flows between these entities.
We denote $K$ as the set of all product types within the system. 
\begin{table}[tb]
\caption{A Notation Summary for the centralized model in \eqref{mip}}
\label{tab:notation}
    \begin{tabularx}{\columnwidth}{lX}
    \hline
    \multicolumn{2}{l}{\textbf{Known Parameters}}\\
    \hline
    $d_{ik}$& demand of product $k$ at vertex $i$; by convention, $d_{ik} < 0$. \\
    $f_{ij}$& fixed transportation cost from vertex $i$ to $j$.\\ 
    $c_{ijk}$& unit transportation cost of flowing product $k$ from vertex $i$ to $j$.\\ 
    $q_{ij}$& mixed-flow capacity of edge $(i, j)$. \\ 
    $\Bar{p}_{i}$& mixed-product production capacity available at vertex $i$.\\
    $e_{ik}$& production cost per unit  product $k$ at vertex $i$.\\ 
    $r_{kk'}$& conversion rate from product $k$ to $k'$, i.e. the unit of product $k$ consumed to produce one unit successor product $k'$.\\ 
    $\phi_{i}$& fixed cost of opening the production line of vertex $i$. \\
    $I_{ik}^0$& initial inventory of product $k$ at vertex $i$ at start of a time period.  \\
    $h_{ik}$& unit holding cost of product $k$ at vertex $i$.\\ 
    $\rho_{ik}^d$& penalty per unit of unsatisfactory of demand $d_{ik}$.
    \vspace{0.3em}
    \\
    \hline
    \multicolumn{2}{l}{\textbf{Decision Variables}}\\
    \hline
    $y_{ijk}$& units of product $k$ flowing from vertex $i$ to $j$.   \\
    $\beta_{ij}$& binary variable, equal to 1 if edge $(i, j) \in E$ is used to transport products, and 0 otherwise.\\
    $x_{ik}$& actual satisfied demand of product $k$ at vertex $i$.\\ 
    $p_{ik}$& units of product $k$ produced at vertex $i$.\\
    $\zeta_{i}$& binary variable, equal to 1 if the production/assembly line opens at vertex $i$, and 0 otherwise. \\
    $I_{ik}$& inventory of product $k$ at vertex $i$ by the end of time period.\\ 
    $\Delta_{ik}^d$& units of unsatisfied demand of product $k$ at vertex $i$.
    \vspace{0.3em}
    \\
    \hline
    \end{tabularx}
\end{table}
Table~\ref{tab:notation} summarizes the parameters and decision variables used in this paper. \\
We compute the total cost of operating the supply chain in Eq.~\eqref{eq:obj}, where $y, x, I, p, \beta, \zeta, \Delta$ indicate the vector forms of the corresponding decision variables shown in Table~\ref{tab:notation}:
\begin{subequations}\label{eq:obj}
\begin{align}
\mathcal{J}&(y, x, I, p, \beta, \zeta, \Delta)\\
=&\sum_{(i,j) \in E, k \in K} c_{ijk}y_{ijk} +
\sum_{i \in V, k \in K}h_{ik} I_{ik} +
\sum_{i \in V, k \in K} e_{ik}p_{ik} \label{obj:var}\\
&+\sum_{(i,j) \in E} f_{ij}\beta_{ij}+
\sum_{i \in V} \phi_{i} \zeta_{i} +
\sum_{i \in V, k \in K}\rho_{ik}^d \Delta_{ik}^d, \label{obj:fix}
\end{align}
\end{subequations}
where \eqref{obj:var} are costs incurred by transporting, manufacturing, and holding products, respectively; \eqref{obj:fix} are fixed costs related to setting up transportation and manufacturing, as well as penalty costs of unsatisfied demand.
We present the centralized model to compute optimal operations in~\eqref{mip}:
\begin{subequations}\label{mip}
\begin{align}
\label{obj:mip}
\min_{y, x, I, p, \beta, \zeta, \Delta}& \hspace{1ex}\mathcal{J}\\
\text{s.t.}\qquad
&\label{cst:flowbalance}
\sum_{j: (i, j)\in E}y_{ijk} - \sum_{j: (j, i)\in E}y_{jik} +  \sum_{k' \in K}r_{kk'}p_{ik'}  \nonumber\\
& \hspace{0ex}- p_{ik} = x_{ik} + I_{ik}^0 - I_{ik}, \ \forall i \in V,\ k \in K \\
\label{cst:cap_y}
&\sum_{k \in K}y_{ijk} \leq q_{ij}\beta_{ij}, \ \forall (i,j) \in E\\
  \label{cst:cap_p}
& \sum_{k \in K}p_{ik} \leq \bar{p}_{i}\zeta_{i},\ \forall i \in V\\
\label{cst:pnlty_d}
& \Delta_{ik}^d \geq x_{ik}- d_{ik},\ \forall i \in V,\ k \in K\\
&  y_{ijk}, x_{ik}, I_{ik}, \Delta_{ik}^d \geq 0,\ \zeta_{i}, \beta_{ij} \in \{0, 1\},\ \nonumber\\
\label{cst:domain}
& \hspace{13ex}\forall i \in V,\ (i, j)\in E,\ k \in K,
\end{align}
\end{subequations}
where constraint in~\eqref{cst:flowbalance} requires the flow balance of each product at each vertex; constraint in~\eqref{cst:cap_y} and~\eqref{cst:cap_p} limits the flow on each edge and production at each vertex by its given capacity; constraint in~\eqref{cst:pnlty_d} computes the unsatisfied demands of each product at each vertex; and constraint in~\eqref{cst:domain} specifies the domains of the decision variables.

\subsection{Disruption response using the centralized model}
\label{sec:disruptioncentral}
To analyze the performance of the proposed multi-agent framework, we first incorporate the effects of various disruptions into Model~\eqref{mip}. 
For example, suppose that $E_d \subset E$ is identified to be a set of edges that become unavailable due to disruptions.
Correspondingly, we add the following constraints to Model~\eqref{mip}: 
\begin{equation}\label{cst:edgelost}
    y_{ijk} = 0,\ \forall (i, j) \in E_d,\ k \in K.
\end{equation}

Once the disruption is identified, a centralized decision-maker will re-run the centralized model with updated network structures, parameters, and constraints to determine the re-optimized decisions.
The objective cost in~\eqref{eq:obj} can be extended by adding costs related to the disruption and response, such as communication cost, network flow change, etc.
We define $\Delta_{ij}^E = \lvert \beta_{ij} -\beta_{ij}^0 \rvert$ and $\Delta_i^V = \lvert \zeta_{i} -\zeta_{i}^{0} \rvert$ as the usage status change on edge $(i,\ j)$ and vertex $i$, where $\beta_{ij}^{0}$ and $\zeta_i^0$ are the binary indicators denoting whether the edge $(i,\ j)$ and vertex $i$ are used in the network before the disruption. 
We also incorporate $\rho_{ij}^E$ and $\rho_i^V$ as unit cost parameters for penalizing changes in the edges and vertices. 

Given the example disruption, we update Model~\eqref{mip} and re-optimize the operations using the following formulation:
\begin{subequations}\label{eq:mipreaction}
\begin{align}
\min\quad & \mathcal{J} + \sum_{(i,j) \in E}\rho_{ij}^{E}\Delta_{ij}^E + \sum_{i \in V}\rho_i^{V}\Delta_i^V \label{eq:newobj}\\
\text{s.t.}\quad 
& \eqref{cst:flowbalance}\text{ -- }\eqref{cst:domain},\ \eqref{cst:edgelost} \label{cst:precst}\\
& \Delta_{ij}^E \geq \pm\left(\beta_{ij} -\beta_{ij}^{0}\right),\ \forall (i,j)\in E,\ k \in K \label{cst:pnlty_E}\\
& \Delta_{ij}^V \geq \pm\left(\zeta_i - \zeta_i^0\right),\ \forall i \in V,\ k \in K \label{cst:pnlty_V}\\
& \Delta_{ij}^E, \Delta_i^V \in \{0, 1\},\ \forall i \in V,\ (i, j) \in E.\label{cst:binary}
\end{align}
\end{subequations}
The added terms in~\eqref{eq:newobj} represent the total penalty costs of changing the used edges and vertices.
In practice, these penalties can represent liquidated damage charges in contracts with local suppliers or transportation companies, or signing fees for new contracts.
Constraints~\eqref{cst:pnlty_E} and~\eqref{cst:pnlty_V} linearize the definitions of $\Delta_{ij}^E$ and $\Delta_i^V$, and constraint~\eqref{cst:binary} specifies their binary value domains.



The generic MILP model is flexible enough to capture different disruptions and incorporate disruption responses by modifying objectives and constraints. 
With the visibility of all entities and their status after the disruptions, this model can provide a globally optimal solution.
In this work, this centralized model serves as a benchmark for evaluating the performance of the proposed multi-agent framework.

\section{multi-agent framework}
\label{sec:multiagentframework}

In this section, we describe the agent network consisting of different types of agents. 
We also provide a detailed design of the model-based agent knowledge base, including capabilities, environments, and state dynamics.
\begin{figure}[tb]
\centerline{\includegraphics[width=1\columnwidth]{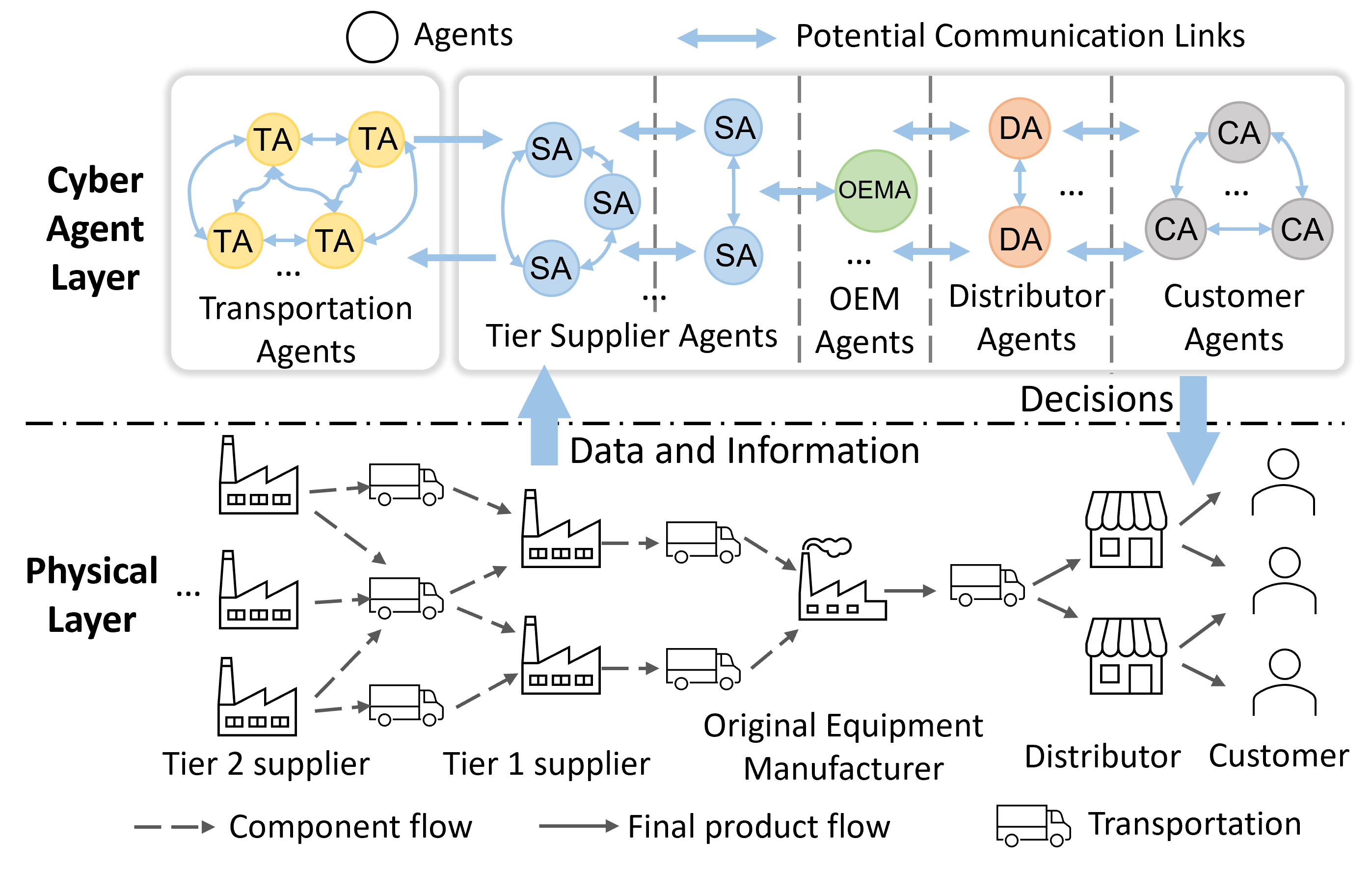}}
\caption{Agents in an example supply network. Each agent represents an associated physical entity. Blue
arrows represent potential communication.}
\label{fig:framework}
\end{figure}

\subsection{Multi-agent supply chain network}
\label{sec:agentnetwork}
Various types of agents have been developed in the existing literature for locally operating a supply chain~\cite{ghadimi2018multi,de2019multi}.
We summarize the most commonly used agent types and use the following types to build a supply chain agent network:
\begin{itemize}
    \item Customer agent - orders products based on demand
    \item Distributor agent - receives the customer’s order and sells products to customers
    \item Original equipment manufacturer (OEM) agent - assembles products using components from tier suppliers
    \item Tier supplier agent - supplies components to the OEMs
    \item Transportation agent - delivers components and products between physical entities
\end{itemize}

Each agent represents a physical entity and will be initialized with the physical entity characteristics, such as capabilities, goals, and risk tolerance.
The agent obtains data from the physical world and is able to command changes to its corresponding physical entity, as shown in~Fig.\ref{fig:framework}.

We generate the agent network based on the physical network graph $G(V,E)$ in Section~\ref{sec:centralizedmodel}.
Here, both the vertices $V$ and edges $E$ are a set $N$ of nodes, which are connected by a set $L_t$ of communication links.
During each time period $t$, we denote the agent network as $G^a_t(N,L_t)$, where $N=V\cup E$.
Since $G$ is assumed to include all potential physical entities, $N$ is static while $L_t$ varies over time.
For a specific agent $A_i$, $G^a_{i,t}(N_i,L_{i,t})$ represents its local communication network at time period $t$, where $N_i \subseteq N$ is a set of nodes that $A_i$ can communicate with via links in $L_{i,t}$.
In this work, each time index $t$ represents one time period where no disruptions occur and the network remains constant, such as flow rates.
Once a disruption occurs, the response action requires changes to the network, and the supply chain enters the next time period $t+1$.

\subsection{Agent knowledge base}
\label{sec:agentknowledgebase}

The proposed agent knowledge base consists of models for agent capabilities, environments, and states.
These models describe the agent knowledge at time $t$.
\subsubsection{Capability model}
The capability model describes the behaviors that an agent can perform in the supply chain network.
Based on the agent classification in Section~\ref{sec:agentnetwork}, agent capabilities are classified using three types: production, inventory, transportation.
The capability model consists of capability knowledge and several associated mapping functions.
We define the capability knowledge of agent $A_i$ as a set $M_{i,t}=\{Pd_{i,t},Iv_{i,t},Tp_{i,t}\}$, where $Pd_{i,t},Iv_{i,t},Tp_{i,t}\subseteq K$ define production, inventory, and transportation sets with which agent $A_i$ can operate.
For example, a product type $k\in Pd_{i,t}$ indicates that $A_i$ can produce product $k$ at time $t$.
Along with these high-level representations of the capabilities, several mapping functions are used to describe the characteristics of a capability, such as cost and capacity.
Table~\ref{tab:capabilitymapping} presents some example mapping functions.

The capability model is built when the agent is initialized based on its associated physical entity.
This information is dynamically updated as the supply chain environment changes, such as cost increases, production line changes, etc.

\begin{table}[tb]
\caption{Examples of mapping functions to describe capabilities}
\begin{center}
\begin{tabular}{cm{20em}}
\hline
\textbf{Agent capability}& \textbf{Example mapping function}\\
\hline
DA: inventory & $Iv_{i,t}\rightarrow \mathbb{R}_+$: mapping from a product type to the cost to hold unit product in inventory.\\
\hline
OEMA: production & $Pd_{i,t}\rightarrow K\times\mathbb{R}_+$: mapping from a product type to the type and amount of needed components to produce unit product.\\
\hline
TA: transportation & $E\rightarrow \mathbb{R}_+$: mapping from an edge to the capacity limit of transporting products.\\
\hline
\end{tabular}
\label{tab:capabilitymapping}
\end{center}
\end{table}
\subsubsection{Environment model}
An agent's knowledge of other agents in the network is encapsulated in the environment model.
As defined in Section~\ref{sec:agentnetwork}, $G^a_{i,t}(N_i,L_{i,t})$ represents the communication network of agent $A_i$ at time period $t$.
Note that $N_i\setminus A_i$ is the set of agents that $A_i$ can communicate with.
In the environment model, these agents are grouped into several sets based on their relationship with agent $A_i$.
In this way, an agent is capable of identifying the subset of agents it needs to communicate and exchange information within the network.
These relationships are modeled as the following mapping functions:
\begin{itemize}
    \item $U_{i,t}:K\rightarrow 2^N$: mapping from product types to upstream agents from which $A_i$ can obtain the products
    \item $D_{i,t}:K\rightarrow 2^N$: mapping from product types to downstream agents to which $A_i$ can provide the products
    \item $T_{i,t}:K\rightarrow 2^N$: mapping from a product type to the transportation agents that can reach agent $A_i$.
    \item $S_{i,t}:M_{i,t}\rightarrow 2^N$: mapping from a capability in agent $A_i$ to the agents that have the same capability 
\end{itemize}
The mapping functions $U_{i,t}, D_{i,t}$, and $T_{i,t}$ identify the agents that can execute a physical product flow (e.g., obtaining products from agent $A_i$) with agent $A_i$.
The information about the flow (e.g., product type and amount) is stored in the environment model via other mapping functions associated with the agent sets.
For example, a function $U_{i,t}(K)\rightarrow K\times \mathbb{R}_+$ maps the upstream agent to the product type and amount of product flow with $A_i$.
The mapping $S_{i,t}$ identifies agents that perform the same behavior as agent $A_i$ for replanning purposes. 

The environment model is generated when the agent is initialized based on its associated physical entity.
The agent sets and their associated information are dynamically updated if the supply chain environment changes, such as production capabilities of a given agent change.
Note that changes may occur during a time period $t$ if the changes only occur in agent knowledge without leading to changes in the network.
\begin{figure}[tb]
\centerline{\includegraphics[width=0.9\columnwidth]{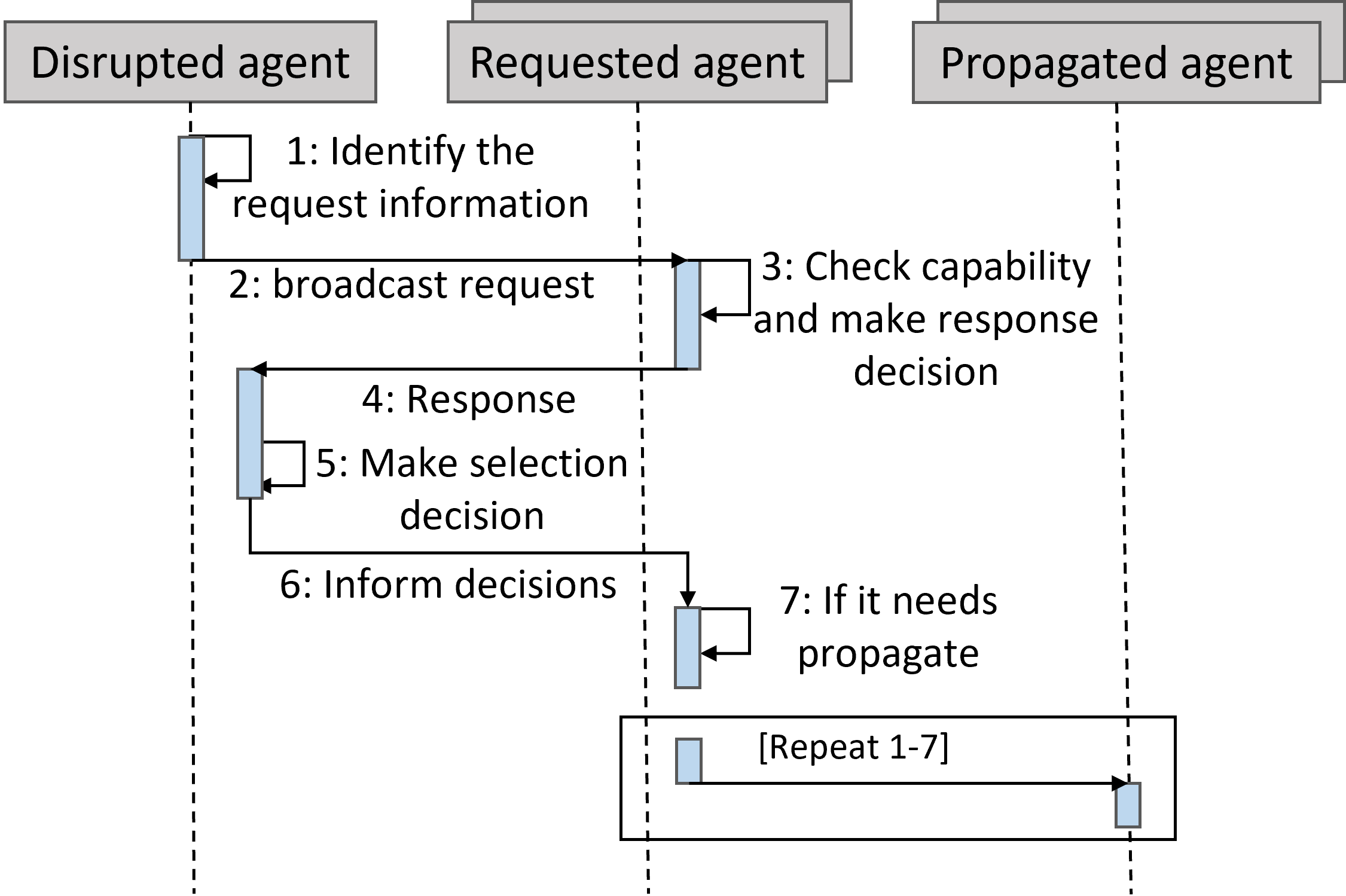}}
\caption{Agent communication scheme to adapt to disruption.}
\label{fig:communication}
\end{figure}
\subsubsection{State model}
The agent dynamics are described as a flow balance of varying input and output products.
The state model describes the dynamics in terms of flow, production, and inventory based on the flow balance equation:
\begin{equation}
    I_{i,t+1} = I_{i,t} + u_{i,t} - z_{i,t} + h_{i,t}(I_{i,t},u_{i,t}).
\label{eq:statedynamics}
\end{equation}
Note that each variable in the state model is a vector indexed by product type, denoted by $[\cdot]_{k\in K}$, where state vector $I_{i,t}=\left[I_{ik,t}\right]_{k\in Iv_{i,t}}$ represents the amount of products held in the agent at time $t$; input vector $u_{i,t}=\left[\sum_{j\in U_{i,t}(k)}y_{jik,t}\right]_{k\in Iv_{i,t}}$ represents product flows into the agent from the upstream nodes $U_{i,t}(K)$; output vector $z_{i,t}=\left[\sum_{j\in D_{i,t}(k)}y_{ijk,t}\right]_{k\in Iv_{i,t}}$ represents the product flows out of the agent to the downstream nodes $D_{i,t}(K)$; and production function $h_{i,t}(I_{i,t},u_{i,t})$ defines the amount of components used to generated new products if the agent has production capabilities.
Note that the variables in~\eqref{eq:statedynamics} are bounded by agent specific limits.
For example, $I_{i,t}$ is limited by the inventory capacity of $A_i$, while $u_{i,t}$ and $z_{i,t}$ are limited by transportation capacities.

\section{agent communication for disruption response}
\label{sec:disruptionresponse}

In this section, we describe the proposed agent communication scheme presented in Fig.~\ref{fig:communication}. 
We assume that agents can communicate no matter how the disruptions affect the physical entities, and a communication link could be established or removed based on the disruption.
The agent communication for disruption response is assumed to be completed within the time period $t$ when the disruption occurs. 
Future work will relax these assumptions. 
For simplicity, the time index is neglected in this section.

\subsection{Agent communication}
\label{sec:agentcommunication}

\subsubsection{Request}
In this work, we assume that agents are able to detect disruptions that occur in their associated physical entities and identify the effects of the disruption by periodically obtaining data and information from the physical entities.
Disruptions can affect the flow ($u,z$), inventory ($I$), and/or production capabilities ($h(I,u)$) of an agent. Therefore, when a disruption occurs, the disrupted agent $(A_d)$ must determine: 1) the product flows needed to recover the flow balance (e.g. satisfy downstream demand); 2) the agents with whom it should communicate for replanning. 
Note that if multiple disruptions occur or multiple agents are disrupted, each agent will send their requests separately; the coupling relationships between disrupted agents is outside the scope of this work.

Consider a case when an OEM loses all of the inventory and production equipment due to an unexpected catastrophic event. The request, $Req=(y_d)$, is defined as $y_d = \left[y_{djk}\right]_{j\in D_d(k),\ k\in Iv_d.}$
In this request, $y_d$ is a vector indexed by downstream agents.
Each element $y_{djk}$ represents the flow of product $k$ from the disrupted OEM $d$ to its downstream agent $A_j$.
Note that the request could include other information related to the flow, such as budget, delivery due date, etc., as additional requirements.
This type of request from OEM $d$ is only sent to the agents that have the same production capabilities denoted by ($S_d(M_d)$) to find alternative OEM agents that can provide the required products to the distributors. 

\subsubsection{Response}
Agents that receive the request will then check their knowledge base and determine their maximum ability to recover the product flows.
In the example of an OEM being unavailable, a response will include available product flow $\bar{y}_{ijk}$ of product $k$ that the requested agent $A_i$ can provide to $A_j$.
Agent $A_i$'s response to OEM $d$ is $\bar{y}_i=[\bar{y}_{ijk}]_{j\in D_d(k),\ k\in Iv_d}$.
Assuming flows are primarily limited by the transportation capacity, the requested agent $A_i$ can determine their response by solving the optimization problem in~\eqref{eq:maxout}:\begin{subequations}
\label{eq:maxout}
\begin{align}
\min_{\bar{y}_i}\quad & \|\bar{y}_i-y_d\|_1 \label{eq:maxoutobj}\\
\text{s.t.}\quad 
&\sum_{k \in Iv_d}\bar{y}_{ijk} \leq q_{ij}, \ \forall j \in D_d(k), \ k\in Iv_d, \label{eq:maxoutcap}
\end{align}
\end{subequations}
where~\eqref{eq:maxoutobj} minimizes the difference between the request and response.
Note that multiple agents can be used to provide the needed flow, and the objective and constraints can be extended if considering cost, production capacity, etc.

\subsubsection{Inform}
After receiving responses from all of the requested agents, the disrupted agent $A_d$ determines the alternative flows $y_r$ that can replace $y_d$: $y_r = \left[\ \sum_{i\in P} y_{ijk}\ \right]_{j\in D_d(k),\ k\in Iv_d} ,$
where $i$ represents the index of the requested agent $A_i$ and $P$ is the set of requested agents.
Note that $y_r$ is a vector indexed by downstream agents.
Assuming no other constraints, the disrupted agent can determine the alternative flow $y_r$ by solving the optimization problem in~\eqref{eq:distribution}:
\begin{subequations}\label{eq:distribution}
\begin{align}
\min_{y_r}\quad & C(y_r)+ \|y_r-y_d\|_1 + \|y_r-y_0\|_1 \label{eq:distributionobj}\\
\text{s.t.}\quad 
&y_{ijk} \leq \bar{y}_{ijk}, \ \forall k \in Iv_d, j \in D_d(k) \label{eq:distributioncap1}\\
&y_r \leq y_d, \label{eq:distributioncap2}
\end{align}
\end{subequations}
where $C(y_r)$ represents the total cost of choosing $y_r$ and $y_0$ represents the product flow vector before the disruption with the same indices as $y_r$.
The objective~\eqref{eq:distributionobj} is to minimize the cost, the unsatisfied total amount of flows, and the flow changes.
The two constraints~\eqref{eq:distributioncap1} and~\eqref{eq:distributioncap2} denote that alternative flows cannot exceed their limitations determined by~\eqref{eq:maxout} and the request.
Note that the objective can be extended to consider delivery time, product quality, etc.
Other constraints, such as budget, delivery due date, etc., can also be added.

\subsubsection{Propagation}
After an agent is selected and informed, it will check whether additional products, such as components from an upstream agent, are needed to provide the new product flow.
If so, the selected agent will send a request to the upstream agents.
The communication follows the process detailed above, as shown in Fig.~\ref{fig:communication}.
It is possible that a solution may not be found due to the limited information of the agents. In such cases, the disrupted agent will request the centralized model to generate a new product flow plan using a global perspective.

\subsection{Network update}
\label{sec:networkupdate}
The agent knowledge base and communication network are updated dynamically based on changes within the network.
For example, new product flow lines must be updated within the relevant agent knowledge base. 
For the capability model, the agents update the capability knowledge set $M_i$ if a capability change occurs, such as changing the production line to produce a new product.
The associated mapping functions are also updated if the characteristics of the capabilities change, such as increased production costs.
For the environment model, communication groups are updated if the agents obtain new knowledge of upstream agents, downstream agents, and agents that have the same capability.
The state model is updated based on the new flow balance of the agent.

\section{Case study}
\label{sec:casestudy}

To test the feasibility and performance of the proposed framework, we implemented the multi-agent framework and agent communication infrastructure in a simulated supply chain network.
In this section, we describe the set-up and simulation results of the supply chain case study.

\subsection{Case study set-up}
\label{sec:setup}

\begin{figure}[tb]
\centerline{\includegraphics[width=0.95\columnwidth]{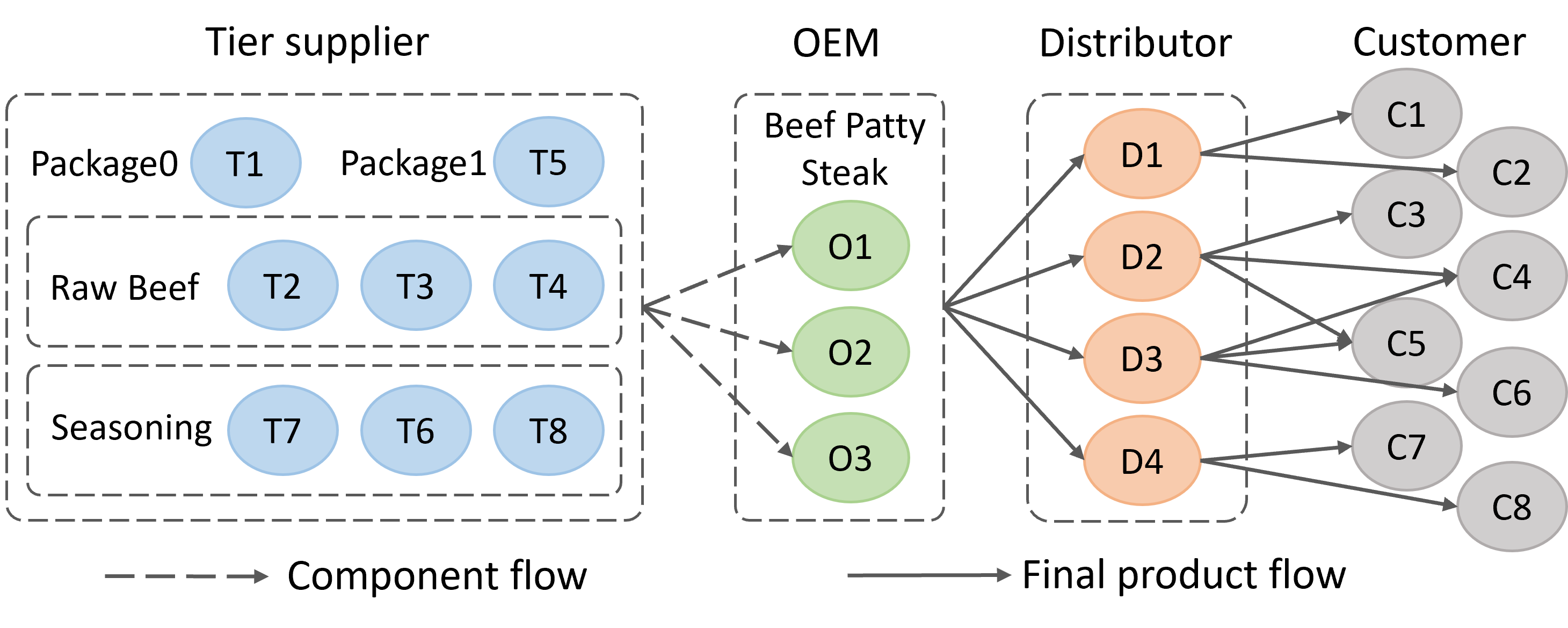}}
\caption{Supply chain network used for case study. Arrows represent component and  product flow via transportation.}
\label{fig:setup}
\end{figure}

The supply chain network we use in this work is an adapted version from reference~\cite{willems2008data}.
The simulated network contains 8 tier suppliers, 3 OEMs, 4 distributors, and 8 customers that are connected via distinct transportation, as shown in Fig.~\ref{fig:setup}.
The annotations represent the product types that the tier suppliers and OEMs can produce and hold in inventory.
For example, the annotation for OEM indicates that all OEMs can produce and hold Beef Patties and Steak.
The distributors have inventory capabilities for both the Beef Patties and Steak.
The customers provide the demand for the Beef Patties or Steak.
Each unit of Beef Patty needs 1 unit of Raw Beef, Seasoning, and Package0, and each unit of Steak needs 1 unit of Raw Beef, Seasoning, and Package1.
Production and transportation in the network correspond to a cost and capacity constraint. Each entity in the network corresponds to an agent that follow the framework outlined in this work and illustrated in Fig.~\ref{fig:setup}. 

\subsection{Case study results}
\label{sec:results}
\begin{figure*}[tb]
\centering
\subfigure[Disruption at C5.]{\includegraphics[width=0.55\columnwidth]{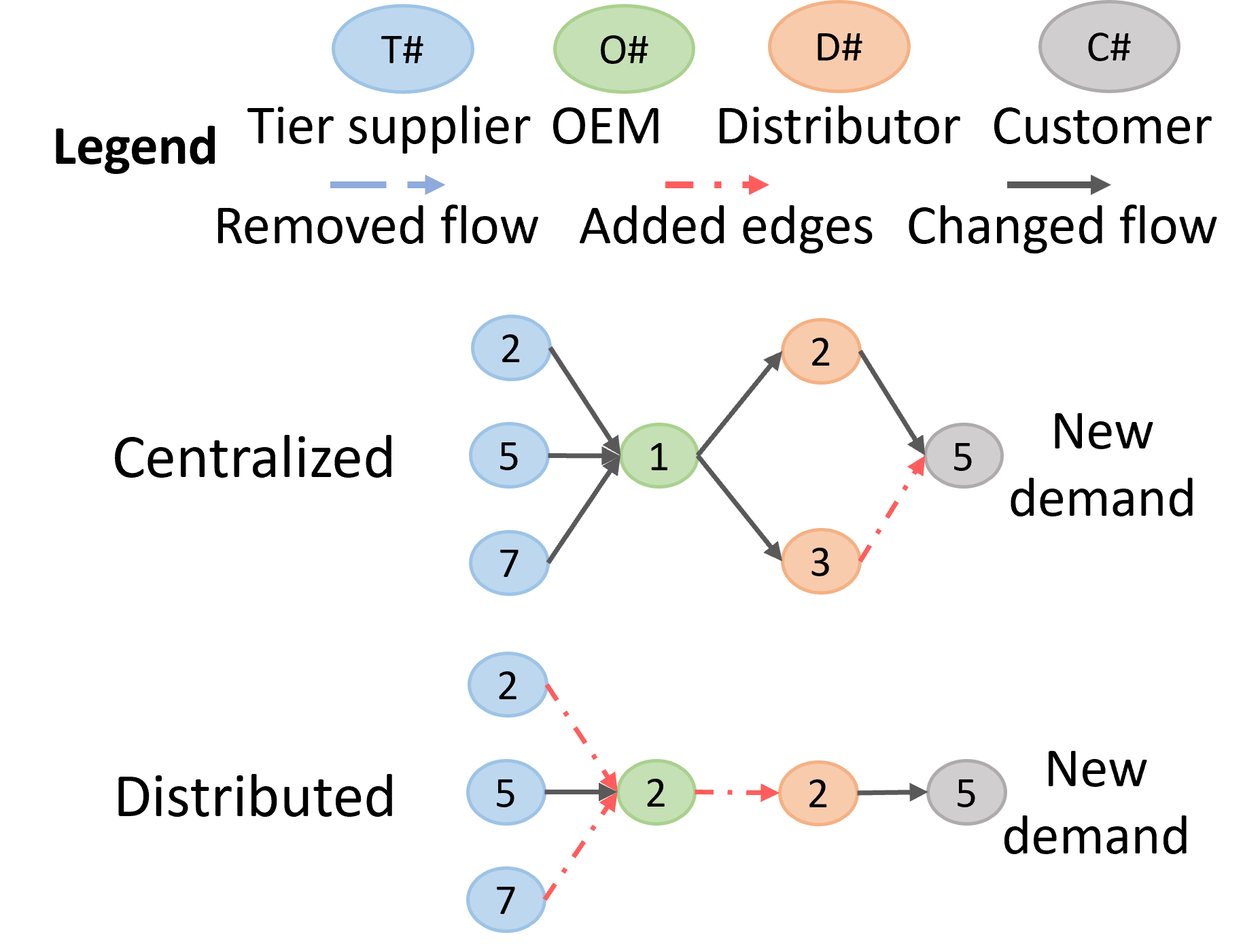}%
\label{fig:C5}}
\hfil
\subfigure[Disruption at  T4.]{\includegraphics[width=0.4\columnwidth]{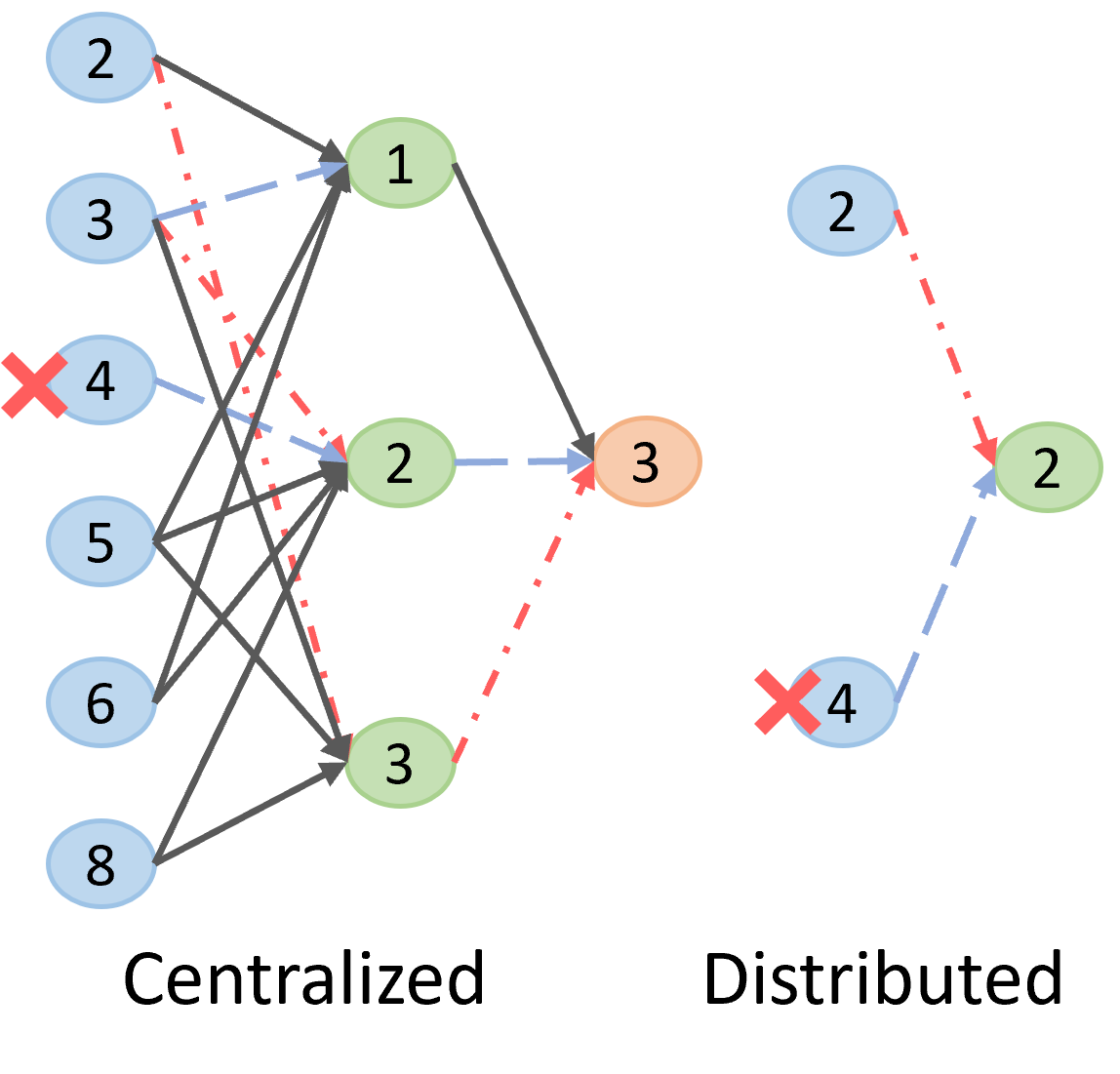}%
\label{fig:T4}}
\hfil
\subfigure[Disruption at O1.]{\includegraphics[width=0.55\columnwidth]{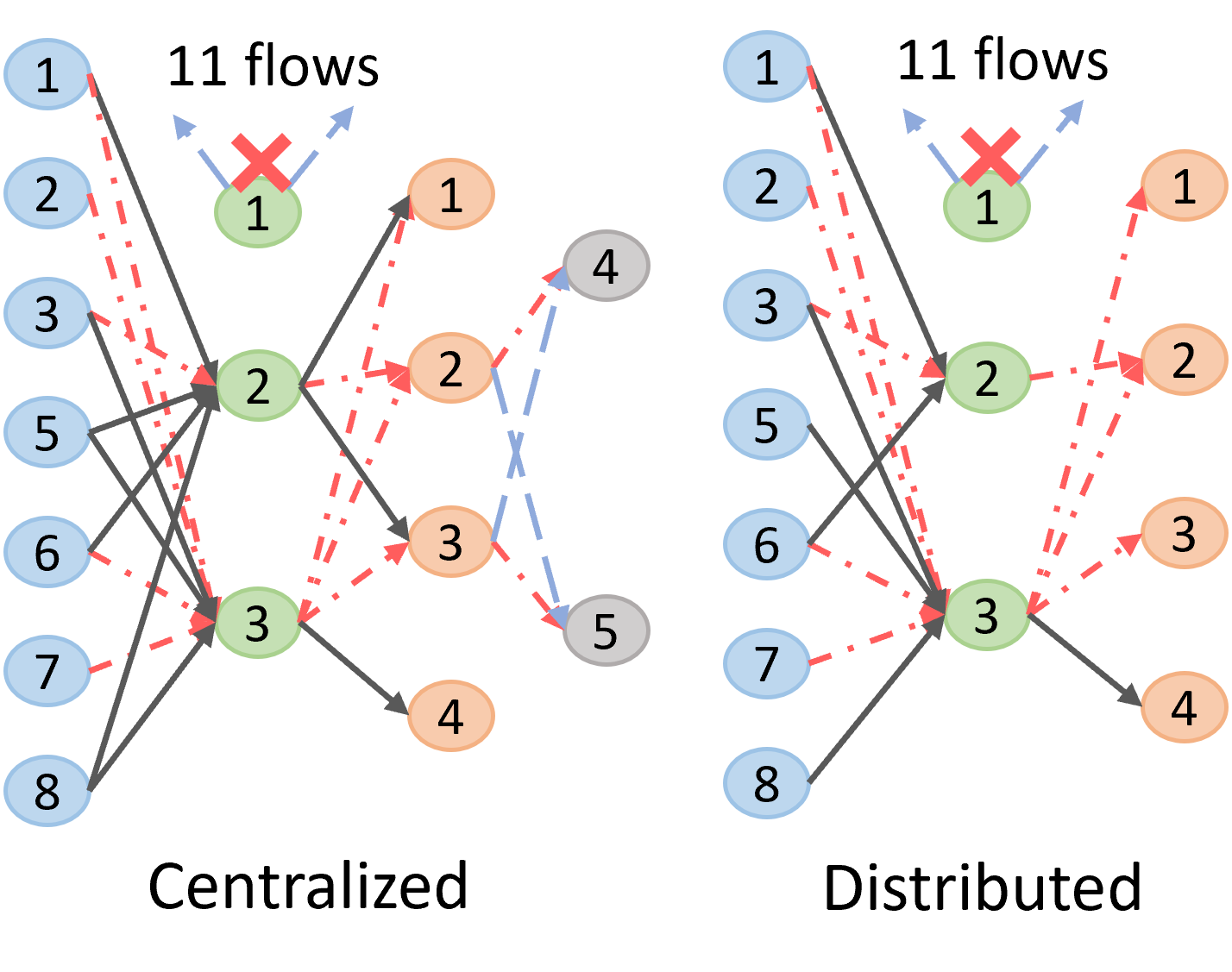}%
\label{fig:O1}}
\hfil
\subfigure[Disruption at O1 with response constraints.]{\includegraphics[width=0.5\columnwidth]{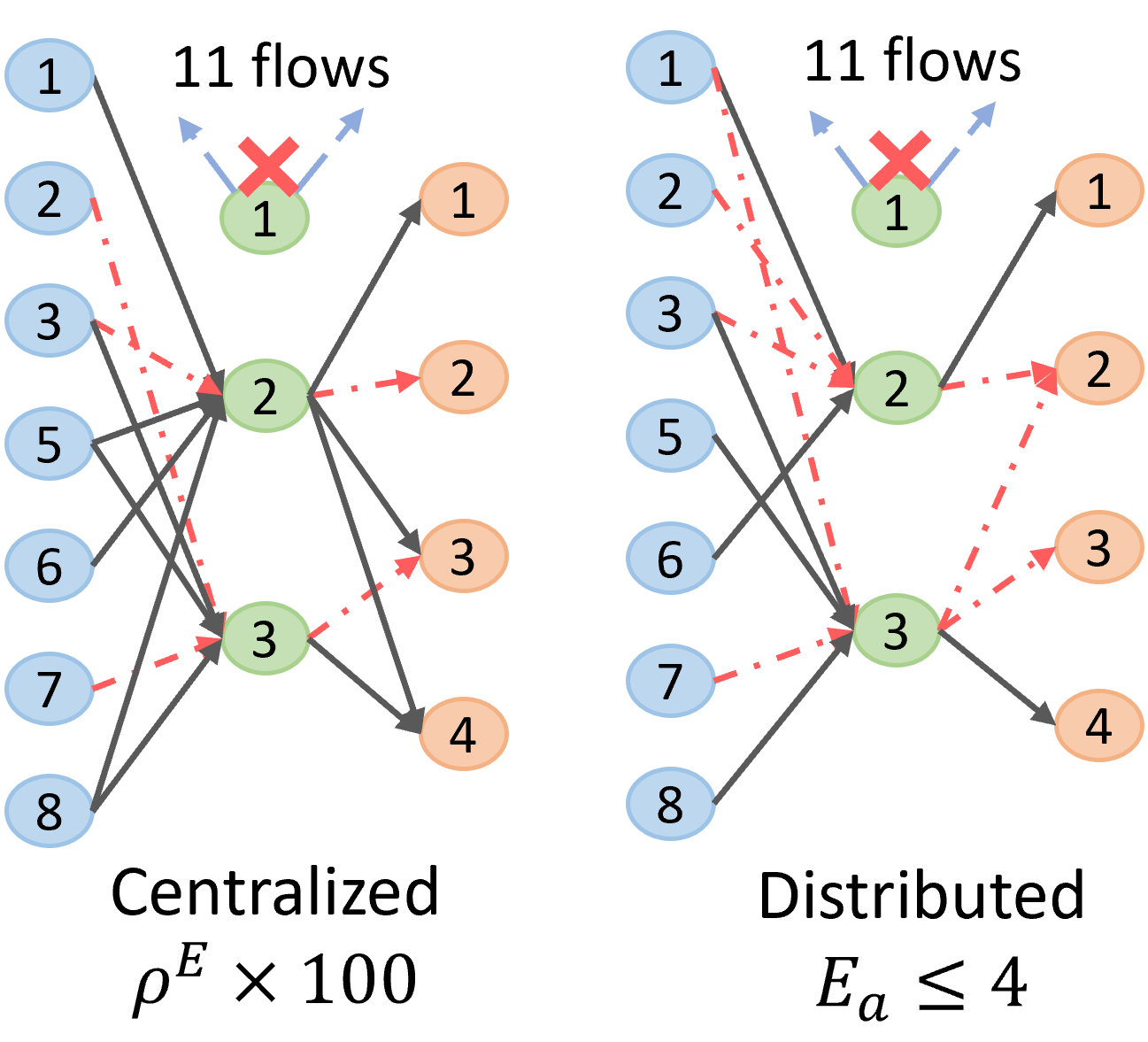}%
\label{fig:O1extend}}
\caption{Changes to product flow in the network by solving the centralized and distributed models to respond to different disruptions.}
\label{fig:flowchange}
\end{figure*}
In this case study, the supply chain starts with an existing product flow plan determine by solving the centralized model~\eqref{mip}.
Based on the network topology, we chose three disruptions to evaluate the performance differences between the more general centralized method and our proposed distributed method: 1) C5 receives a new demand request; 2) T4 becomes unavailable; and 3) O1 becomes unavailable.
Note that C5 is a downstream agent with only inflow, T4 is an upstream agent with only outflow, and O1 is an agent with both inflow and outflow.
The performance of the framework based on the different disruption scenarios is evaluated using the following metrics:
\begin{itemize}
    \item Flow cost change $(C_f)$: change in the costs associated with a given flow 
    \item Production cost change $(C_p)$: change in the production costs 
    \item Added edges $(E_a)$: number of edges that are added to address the disruption
    \item Changed flows $(F_c)$: number of flow channels that are changed in terms of type and/or amount of product
    \item Communication effort $(C_e)$: number of communication exchanges used to derive a response to the disruptions
\end{itemize}
\begin{table}[tb]
\caption{The performance of centralized and proposed distributed method under different disruptions}
\centering
\begin{tabularx}{\columnwidth}{Xcccccc}
\hline
\multicolumn{2}{c}{\textbf{Scenarios}}& \multicolumn{5}{c}{\textbf{Metrics}}\vspace{0.2em}\\
Disrupted agent& Methods & $C_f$ & $C_p$ & $E_a$ & $F_c$ & $C_e$\vspace{0.3em}\\
\hline 
\noalign{\vskip 0.3em}
\multirow{2}{1em}{C5}& Centralized &+1599.54 &+5691& 1 & 6 & 54\\
& Distributed &+1722.39 &+5657.89& 3& 2& 29\vspace{0.3em}\\
\hline
\noalign{\vskip 0.3em}
\multirow{2}{1em}{T4}& Centralized &-1742.72 &+5889.56 & 3 & 10 & 56 \vspace{0.2em}\\
& Distributed &-1361.46 &+6039.89 &1 & 0 & 6\vspace{0.3em}\\
\hline
\noalign{\vskip 0.3em}
\multirow{5}{1em}{O1}& Centralized &+1981.49 &-596.31& 11 &10& 61 \vspace{0.2em}\\
& Distributed &+2215.31&-246.9& 9& 6 &46 \vspace{0.2em}\\
& Centr. ($\rho^E\times100$) &+2154.78&-625.31& 5&11&61 \vspace{0.2em}\\
& Centr. ($E_a\leq4$) &\multicolumn{5}{c}{unsatisfying demand} \vspace{0.2em}\\
& Distr. ($E_a\leq4$) &+2309.86&-201.91 & 7 & 7 & 46 \vspace{0.3em}\\
\hline
\end{tabularx}
\label{tab:results}
\end{table}

The calculations of $C_f, C_p, E_a$ and $F_c$ are straightforward by definition.
The communications effort $C_e$ in the centralized method includes the request for re-running the model, the requests to and responses from all the nodes in the supply chain to collect information, and the notifications to the nodes whose flow and/or production plan need to change.
In the distributed method, $C_e$ includes all agent requests, responses, and inform messages, as defined in Section~\ref{sec:agentcommunication}.

Table~\ref{tab:results} shows the performance of the centralized and proposed distributed methods under different disruptions. 
Figure~\ref{fig:flowchange} represents the changes in product flow within the network.
In the scenario with a disruption at C5, the centralized method re-distributes the product flow to avoid adding edges, which leads to more overall changes to the existing flow edges, as shown in Fig.~\ref{fig:C5}. 
The proposed distributed method minimizes undesirable local changes without increasing the cost drastically.
However, the distributed approach adds more new edges since the local agents cannot provide knowledge of the added edges of other agents. 
For example, in the simulation, agent D2 does not know that O2 needs new edges from T2 and T7 to satisfy its request.
The results also show that new demands from customers require minimal flow changes from downstream to upstream.

The scenario of a disruption at T4 illustrates the benefit of using the proposed multi-agent method to respond to disruptions at an upstream node.
Although the flow and production costs are higher than the centralized method, the agent communication only occurs within the agents that have the same capability as T4 and related OEMs, which results in a faster response.
Fig.~\ref{fig:T4} shows the distributed method only results in a small change in the network.

In the scenario of a disruption at O1, both the centralized and distributed methods lead to lots of edge and flow changes with a large communication effort, as shown in Fig.~\ref{fig:O1}.
Since O1 has a combined 11 inflow and outflow edges before the disruption, the distributed method cannot recover the complete flow demand within a local communication environment.
In practice, adding new flow edges can lead to high costs, thus we increase the penalty $\rho^E$ on the addition of new edges and add a constraint to $E_a$ to check the performance.
Figure~\ref{fig:O1extend} shows that the centralized method provides a solution with a smaller $E_a$ when a larger $\rho^E$ is applied, while the cost becomes higher.
With the hard constraint on $E_a$, the centralized model fails to provide a solution that can satisfy the customer demand.
However, in the distributed method, since agents only have their local view, the value of $E_a$ only considers their connected edges. Thus, the distributed method can provide a feasible solution, as shown in Fig.~\ref{fig:O1extend}.

Overall, both the centralized and distributed methods lead to higher costs to satisfy the demand.
The centralized method provides lower-cost solutions since it has a global view of the network. 
However, it often leads to more production and flow changes with larger communication efforts.
The distributed method responds to the disruptions with less communication and maintains the original flow as much as possible, especially under a small disruption. 
The results from network changes also show that the proposed multi-agent method performs better when the disruption occurs in an upstream node or a node with few initial flows.
An enterprise can make a decision depending on whether they aim for a quick response, a low-cost solution, or a minimal-change solution.

\subsection{Insight from case studies}
\label{sec:insight}

The case study presented in this section showcases the feasibility of using a model-based multi-agent framework to drive the behavior of the entities in the supply chain network.
In the case study, both centralized and distributed models seek to minimize the number of edge changes for decision-making.
However, the objective functions can include additional information such as desired inventory level or communication costs.

Additionally, we initialize the local view of the agents with a subset of the global supply chain network in the case study.
In practice, however, an agent may have knowledge of some entities outside of the network or the centralized model.
Incorporating this knowledge and designing an agent communication strategy for a new entity joining the network can be studied in future work.

The case study illustrates that disruptions at different nodes lead to different performance outcome.
We will investigate nodes' attributes that result in these differences in future work.
Future work may also include the analysis of the effects of network topology on response performance, and the design of an optimization framework for selecting response strategies.

\section{conclusion}
\label{sec:conclusion}

Various multi-agent frameworks have been studied to cope with disruptions in supply chain systems.
In this paper, we introduce a multi-agent framework where agents are equipped with a model-based knowledge base and dynamic communication network.
The proposed agents have the knowledge of their capabilities, environments, and state dynamics.
Based on agent knowledge, this paper develops an effective agent communication strategy to recover a new product flow plan without requiring prior knowledge of the potential disruptions.
The proposed work can be used to create supply chain models that enable dynamic and agile responses to disruptions and allow for greater agent flexibility and scalability to larger systems when compared to rule-based architectures.
Showcased through a simulated case study, the proposed work provides an agile response to supply chain disruptions and maintains the original flow as much as possible.
Future work will include generalizing the agent model and communication infrastructure for different disruptions, analyzing how the different disrupted agents and network topology affect the performance, and designing an optimization framework for determining different strategies to respond to disruptions.

\bibliographystyle{IEEEtran.bst}
\bibliography{refs.bib}

\end{document}